\documentclass[12pt]{article}
\newcommand{\dep}{\partial_+}
\newcommand{\dm}{\partial_-}

\newcommand{\plabel}{\label}

\usepackage{amstex}
\usepackage{cite}

\begin{document}

\begin{titlepage}
\renewcommand{\thefootnote}{\fnsymbol{footnote}}

\hfill TUW--97--07  \\
\begin{center}
\vspace{1cm}

{\Large\bf  Hawking Radiation for Non-minimally Coupled Matter 
from Generalized 2D Black Hole Models}

\vfill
\renewcommand{\baselinestretch}{1}

{\bf W.\ Kummer$^1$\footnotemark[1], H. Liebl$^1$\footnotemark[2]
       and D.V.\ Vassilevich$^{1,2}$\footnotemark[3]}

\vspace{7ex}

{$^1$Institut f\"ur
    Theoretische Physik \\ Technische Universit\"at Wien \\ Wiedner
    Hauptstr.  8--10, A-1040 Wien \\ Austria}

\vspace{2ex}

{$^2$Department of Theoretical Physics \\
   St.\ Petersburg University \\ 198904 St.\ Petersburg \\  Russia}


\footnotetext[1]{e-mail: \texttt{wkummer@@tph.tuwien.ac.at}}
\footnotetext[2]{e-mail: \texttt{liebl@@tph16.tuwien.ac.at}}
\footnotetext[3]{e-mail: \texttt{vasilev@@snoopy.niif.spb.su}}


\end{center}
\vfill

\begin{abstract}
It is well known that spherically symmetric reduction of General 
Relativity (SSG) leads to non-minimally coupled scalar matter. We 
generalize (and correct) recent results to Hawking radiation for 
a class of dilaton models which share with the Schwarzschild black hole
non-minimal coupling of 
scalar fields and the basic global structure. An inherent ambiguity of such 
models (if they differ from SSG) is discussed. However, for SSG we obtain
the rather disquieting result of a {\it negative} Hawking flux at infinity,
if the usual recipe for such calculations is applied. 
\end{abstract}
\end{titlepage}
\vfill

\section{Introduction}

It has been known for a long time [1] that spherical reduction of 
d = 4  General Relativity (GR) for scalar matter leads to 
non-minimal coupling. We fix the radius to be a dimensionless 
quantity in the line element 

\begin{equation}
ds^2=g_{\mu\nu}dx^\mu dx^\nu +e^{-2\phi}d\Omega^2 \quad .
\plabel{ssa}
\end{equation}
The metric $g_{\mu\nu}$ refers to the variables $x^\mu = (x^0,x^1)$. 
With the dilaton field $\phi (x) = -\ln r$ the effective action 
in $d = 1 + 1$ for scalar matter $f$ becomes 

\begin{equation}
S_{m}=-\frac 12 \int d^2x \sqrt{-g} e^{-2\phi}
g^{\mu\nu}\partial_\mu f\partial_\nu f \quad .
\plabel{nmc}
\end{equation}
Covariant derivatives will refer to the metric $g_{\mu\nu}$
in the following.

Somewhat surprisingly until very recent times 
\cite{bousso,noj97a,ElNO,mik97,noj97b} no 
computation of the Hawking radiation for that case seems to 
exist. The purpose of our present note is to give a comprehensive 
and direct answer to that question including all physically 
interesting models which generalize spherically symmetric gravity 
(SSG). This also allows us to improve the results of 
\cite{bousso,noj97a,ElNO,mik97,noj97b} and to 
show the arbitrariness involved when SSG is generalized. Therefore, 
in the following we take the geometric part of our action to be 
\cite{KKL} 

\begin{equation}
\plabel{ldil}
S_{gr}=\int d^2x \sqrt{-g}e^{-2\phi}(R+4a(\nabla\phi)^2 +Be^{4(1-a)\phi}).
\end{equation}

This class of dilaton models covers all asymptotically flat 
theories with one horizon and arbitrary power behavior in Kruskal 
coordinates at the singularity. It contains SSG as the special 
case $a = \frac{1}{2}$. The string-inspired dilaton black hole 
\cite{man91} follows for $a = 1$. For all $a \leq 1$ and $B > 0$ the 
global structure from (\ref{ldil}) corresponds to the Penrose diagram of 
the Schwarzschild black hole, although a 'Schwarzschild gauge' is 
possible for the interval $0 < a \leq 1$ only with the limiting 
case of $a = 1$ \cite{man91} showing the potentially dangerous feature of 
a non-null incomplete but null
complete singularity \cite{KKL}.

\section{Conformal Anomaly}

In the SSG case the (ultralocal) measure for the matter 
integration is well defined, because

\begin{equation}
\plabel{ca1}
\int d\Omega \sqrt{-^4g} =  e^{-2\phi}\sqrt{-g} \quad ,
\end{equation}
where $^4g$ is the determinant of the 4 dimensional metric in GR.
For the generalized class of models (\ref{ldil}), however, this definition 
is not unique, as well as the one for an eventual non-minimal 
factor for the possible coupling to matter in (\ref{nmc}). In 
that case we have to allow the general replacements $\phi \to 
\varphi (\phi)$ in (\ref{nmc}) and $\phi \to \psi (\phi) $ in 
(\ref{ca1}), where 
$\varphi$ and $\psi$ may be general (scalar) functions of the 
dilaton field. 
The measure ${\cal D}f$ is defined by requiring that 
$\int {\cal D}f \exp (i \int d^2x \sqrt{-g} e^{-2\psi (x)} f^2(x))$
is a field independent constant.
With these replacements and in terms of the field 
$\tilde f = f\; e^{-\psi}$ which satisfies the standard 
normalization condition, (\ref{nmc}) can be rewritten as
\begin{equation}
\plabel{tact}
S = -\frac 12 \int \sqrt{-g}d^2x\ 
\tilde f A \tilde f \quad ,
\end{equation}
where
\begin{equation}
A = -e^{-2\varphi +2\psi}g^{\mu\nu} (\nabla_\mu \nabla_\nu
+2(\psi_{,\mu}-\varphi_{,\mu})\nabla_\nu +\psi_{,\mu\nu}
-2\varphi_{,\mu} \psi_{,\nu} )\quad . \plabel{A}
\end{equation}
The path integral for $\tilde f$ leads to the effective action 

\begin{equation}
W = \frac 12  \mbox{Tr} \ln A \quad .
\end{equation}

After continuation to the Euclidean domain $A$ becomes an 
elliptic second order differential operator.
The corresponding one loop effective action $W$ can be expressed 
in terms of the zeta function of the operator $A$ 
\footnote{For a clear and extensive discussion on the $\zeta$ function
technique consult the recent monograph \cite{esposito}.}:
\begin{equation}
W=-\frac 12 \zeta'_A(0), \qquad \zeta_A(s)={\rm Tr}(A^{-s})
\plabel{zeta}
\end{equation}
Prime denotes differentiation with respect to $s$. From $W$ 
regularized in this way an infinitesimal conformal transformation 
$\delta g_{\mu\nu} = \delta k g_{\mu\nu}$ produces the 
trace of the (effective) energy momentum tensor

\begin{equation}
\delta W=\frac 12 \int d^2x \sqrt g\delta g^{\mu\nu}T_{\mu\nu}
=-\frac 12 \int d^2x \sqrt g \delta k(x)T_\mu^\mu (x)
\plabel{T}
\end{equation}

Due to the multiplicative transformation property $\delta A = -\delta k A$ 
of (6) (valid in $d = 2$ only) with the definition of a 
generalized $\zeta$-function

\begin{equation}
\zeta (s|\delta k,A)={\rm Tr}(\delta kA^{-s})
\plabel{varW}
\end{equation}
the variation in (\ref{T}) can be identified with 
\begin{equation}
\plabel{TX}
\delta W=-\frac 12 \zeta (0|\delta k,A) \quad ,
\end{equation}
where we used $\delta \zeta_{A_k}(s)=s \mbox{Tr} (A^{-1} \delta k)$.
Combining (\ref{TX}) and (\ref{T})  we obtain
\begin{equation}
\zeta (0|\delta k,A)=\int d^2x \sqrt g \delta 
k(x)T_\mu^\mu (x)\; .
\plabel{T2}
\end{equation}

By a Mellin transformation one can show that 
$\zeta (0|\delta k,A)=a_1(\delta k,A)$ \cite{gil75}, where $a_1$ is
defined as a coefficient in a small $t$ 
asymptotic expansion of the heat kernel:
\begin{equation}
{\rm Tr}(F\exp (-At)) =\sum_n a_n (F,A)t^{n-1}
\plabel{hk}
\end{equation}
To evaluate $a_1$ we use the standard method \cite{gil75}.
To this end we represent $A$ as
\begin{equation}
A=-(\hat g^{\mu\nu}D_\mu D_\nu +E) ,\qquad 
E=\hat g^{\mu\nu}(-\varphi_{,mu}\varphi_{,\nu}
+\varphi_{,\mu\nu}) \plabel{newA}
\end{equation}
where $\hat g^{\mu\nu}=e^{-2\varphi +2\psi}g^{\mu\nu}$,
$D_\mu =\nabla_\mu +\omega_\mu$, $\omega_\mu =\psi_{,\mu}-
\varphi_{,\mu}$. For $a_1$ follows \cite{gil75}
\begin{equation}
a_1 (\delta k, A)=\frac 1{24\pi} \int d^2x\sqrt{-\hat g} \delta k  
(\hat R+6E)\; .
\plabel{a1}
\end{equation}
Returning to the initial metric and comparing with (\ref{T}) 
we obtain the most general form of the `conformal anomaly' 
\footnote{We use this expression although for a non-Weyl invariant
geometric part (\ref{ldil}) of the action 
is broken already at the classical level.}
for non-minimal coupling in d=2:
\begin{equation}
T_\mu^\mu =\frac 1{24\pi} (R-6(\nabla \varphi )^2 +
4\Box \varphi +2\Box \psi )
\plabel{T3}
\end{equation}

\section{Hawking radiation}

In a background given by (\ref{ldil}) the matter action transforms
simply under conformal transformations which may be used for important 
simplifications.
In  conformal gauge $g^{+-}=-2e^{-2\rho}$ (\ref{T3}) reduces to 
\begin{equation}
T_{+-}=-\frac 1{12\pi} (\dep \dm \rho +3\dep \varphi \dm \varphi
-2\dep \dm \varphi -\dep \dm \psi )\; .
\plabel{Tpm}
\end{equation}

For SSG alone $\varphi^{SSG} = \psi^{SSG} = \phi$ are  
determined uniquely. In order to see the ambiguities involved for a 
'generalized' theory it is sufficient to consider a general linear 
dependence on the dilaton field,  $\varphi =\alpha \phi$,
$\psi =\beta\phi$.

The Hawking flux is expressed in terms of $T_{--}$ which can be
obtained by integrating the conservation condition for the 
energy momentum tensor \cite{chr77} 
\begin{equation}
0=\nabla^\mu T_{\mu -}=
\dep T_{--}+\dm T_{+-}-2(\dm \rho )T_{+-} \quad .
\plabel{cons}
\end{equation}
An integration constant is defined by the condition at the horizon 
\cite{chr77,LVA}:
\begin{equation}
T_{--}\vert_{\rm hor}=0 \plabel{bc}
\end{equation}

We are now able to follow closely \cite{LVA} where the corresponding 
techniques have been used for minimally coupled scalars in 
theories of the type (\ref{ldil}).

Let us first study the CGHS model \cite{man91} with $a = 1$. In this 
particular case the residual gauge freedom of
the conformal gauge can be used to set $\rho =\phi$. 
Then the contribution 
of the last two terms in (\ref{Tpm}) to $T_{--}$ is proportional to that of the 
the first term, which, in turn, is the well known one for minimal 
coupling. Integrating these three terms in (\ref{cons}) with (\ref{bc}) yields 

\begin{equation}
T^{(1)}_{--}\vert_{asymp}=\frac{\lambda^2}{48\pi}(1-2\alpha -\beta)
\plabel{T1-}
\end{equation}
where $4\lambda^2=B$. 
To evaluate the contribution of the second term in (\ref{Tpm}) we need
a slightly different coordinate system \cite{LVA}:
\begin{equation}
ds^2=L(U)(-d\tau^2 +dz^2), \quad dU=L(U)dz
\plabel{newc}
\end{equation}
 For the CGHS model \cite{LVA} we have also
\begin{equation}
L(U)=\frac{e^{\sqrt{B}U}C}{2B}-1\; , \qquad \phi =\sqrt{\frac B4} U 
\end{equation}
where
$C$ is a real parameter proportional to the mass of the black hole. 
The Killing norm L(U) vanishes at the horizon and the  
asymptotically flat region corresponds
to $U\to -\infty$. Eq.\ (\ref{cons}) is now also easily integrated
with (\ref{bc}), and we arrive at the the Hawking
flux in the CGHS model:
\begin{equation}
T^{CGHS}_{--}\vert_{asymp}=\frac{\lambda^2}{48\pi}(1+
\frac 32 \alpha^2-2\alpha -\beta)
\plabel{CGHS}
\end{equation}

Even for minimal coupling ($\alpha = 0$) this expression is 
inherently ambiguous due to the constant $\beta$ which had its 
roots in the ambiguous definition of an ultralocal measure. 
Increasing $\alpha$ (non-minimal coupling) above $\alpha = 4/3$ 
tends to increase $T_{--}$. Of course, by adjusting $\beta$ the 
flux may become zero or even negative as well (`cold dilaton 
black hole'). Like the (geometric) Hawking temperature 
the expression (\ref{CGHS}) for the CGHS model
does not depend on the mass of the black hole. 

For other values of $a$ a simplification by using a residual 
gauge freedom is not possible. Nevertheless, starting from (\ref{newc}) 
with \cite{KKL,LVA}  
\begin{eqnarray}
L(U)&=&\frac {aC}{2B} \left[ \frac {U^2(a-1)^2B}a \right]^{\frac a{2(a-1)}}
-1 \nonumber \\
U&=&\sqrt {\frac aB} \frac {e^{-2(1-a)\phi}}{a-1}
\plabel{UL}
\end{eqnarray}
the integration of (\ref{cons}) is rather elementary, though a bit tedious. 
The final result reads

\begin{equation}
T^{(a)}_{--}\vert_{asymp}=
\frac{1}{48 \pi}T_H^2
\left ( 1-\frac{3\alpha^2}{2(2-a)} -
\frac{1}{2-a} (2\alpha +\beta ) \right ) 
\plabel{Ta}
\end{equation}
which has been expressed in terms of the general geometric Hawking temperature
for the models (\ref{ldil})
\begin{equation}
T_H^2=\frac{a^2}{8} 
C^{\frac{2(a-1)}{a}}\left( \frac{2B}{a} \right)^{\frac{2-a}{a}} \quad .
\end{equation}

Taking the limit $a \to 1$ in (\ref{Ta}) does not reproduce (\ref{T1-})
since there is no smooth limit in the solutions (\ref{UL}). 
The difference in the sign of the second term in the brackets
is due to a finite contribution of $T_{+-}|_{hor}$ for the CGHS model
whereas this term vanishes for all other cases. 
For SSG all parameters are unambiguously defined ($a = \frac 12, \alpha = 
\beta = 1$). Then the bracket in (\ref{Ta}) yields a factor $-2$, i.e. a
negative flux!

\section{Discussion}

The trace anomaly for $T_{\mu \nu}$ in the case of 
non-minimally coupled scalar fields with 
trivial measure ($\psi = 0$) recently has been the subject of several studies 
\cite{bousso,noj97a,ElNO,mik97}. The only difference to our result 
(\ref{T3}) is a 
coefficient in front of $\Box \varphi$. It should be noted that at 
this point these authors also disagree among themselves. The 
reason is that their methods \cite{bousso,noj97a,noj97b} cannot fix total 
derivatives unambiguously. Nojiri and Odintsov define the trace 
anomaly through the equation $\Gamma_{div} \propto \int d^2x 
\sqrt{g} T_\mu^\mu$ with the correct expression \cite{ElNO} for 
the divergent part of the effective action $\Gamma_{div}$. Of 
course, an arbitrary total derivative can be added to $T_\mu^\mu$ 
without changing $\Gamma_{div}$. It is interesting to note that 
complete agreement with our expression (\ref{T3}) is achieved, if 
in \cite{noj97a,noj97b} one 
replaces the expressions for scalar curvature and 
metric by conformally transformed ones (see (\ref{a1})). Bousso and 
Hawking use global scale transformation to relate the coefficients 
in front of $R$ and $\Box \varphi$. This method allows for some 
ambiguity as the authors admit themselves \cite{bousso}. Mikovic and 
Radovanovic calculate the stress energy tensor by varying the 
finite part of the effective action. However, they perform scale 
transformations of quantum fields (see eqs.\ (2.4) and (2.25) in 
\cite{mik97}). In the presence of the conformal anomaly, such 
transformations, in general, change the effective action. Thus 
some contributions can be overlooked in such an approach.
In \cite{ichi} it was suggested that the coefficients in front of 
$\Box \varphi$ are due to boundary terms of the classical action.

The crucial difference of our method is the use of a {\it local} 
scale transformation inside the zeta function. Due to the 
presence of an arbitrary {\it function} $\delta k$ one cannot 
integrate by parts in (\ref{T2}) in order to remove a total derivative 
in (\ref{T}). Hence, all terms there are fixed unambiguously. 

Another attempt to calculate Hawking radiation for non-minimally 
coupled scalar fields was made in \cite{noj97b}. The physical content of 
the model considered there is totally different from ours and 
therefore will not be discussed here. 

Our result for the `anomaly', eq.(\ref{Tpm}), is the most general one
obtainable in 1+1 dimensional theories. 
The result for the Hawking flux in SSG, on the other hand, taken literally
would mean that an influx of matter is necessary to maintain
in a kind of thermodynamical equilibrium the Hawking temperature
of a black hole -- in complete contradiction to established black hole wisdom.
However, to put this result on a sound basis
the treatment of Hawking radiation in the asymptotic region 
in that case certainly requires to go beyond
the usual approach adopted also in our present paper. After all, non-minimally
coupled scalar fields are strongly coupled in the asymptotic region. 
Therefore a result like (\ref{Ta}) for SSG cannot be the final answer. In
fact, probably new methods for extracting the flux towards infinity in such
a case have to be invented.

\section*{Acknowledgement}

This work has been supported by Fonds zur F\"orderung der
wissenschaftlichen For\-schung (FWF) Project No.\ P 10221--PHY.  One
of the authors (D.V.) thanks GRACENAS and
the Russian Foundation for 
Fundamental Research, grant
97-01-01186, for financial support.

\vfil

\end{document}